\documentclass[%
reprint,
aps,
pra
]{revtex4-2}

\usepackage[usenames]{color}
\usepackage{colortbl}
\usepackage{graphicx}
\usepackage{dcolumn}
\usepackage{bm}
\usepackage{amsmath}
\usepackage[utf8]{inputenc}

\usepackage{comment}

\begin{document}

\preprint{APS/123-QED}

\title{Transient Temperature Dynamics of Reservoirs Connected Through an Open Quantum System}

\author{I. V. Vovchenko}
\affiliation{%
 Moscow Institute of Physics and Technology, 9 Institutskiy pereulok, Dolgoprudny 141700, Moscow region, Russia;
}%
\affiliation{%
Kotelnikov Institute of Radioengineering and Electronics, Mokhovaya 11-7, Moscow, 125009, Russia
}%


\author{A. A. Zyablovsky}
\affiliation{
 Dukhov Research Institute of Automatics (VNIIA), 22 Sushchevskaya, Moscow 127055, Russia;
}
\affiliation{
 Moscow Institute of Physics and Technology, 9 Institutskiy pereulok, Dolgoprudny 141700, Moscow region, Russia;
}%
\affiliation{
 Institute for Theoretical and Applied Electromagnetics, 13 Izhorskaya, Moscow 125412, Russia;
}
\affiliation{%
Kotelnikov Institute of Radioengineering and Electronics, Mokhovaya 11-7, Moscow, 125009, Russia
}%

\author{A. A. Pukhov}
\affiliation{
 Moscow Institute of Physics and Technology, 9 Institutskiy pereulok, Dolgoprudny 141700, Moscow region, Russia;
}%
\affiliation{
 Institute for Theoretical and Applied Electromagnetics, 13 Izhorskaya, Moscow 125412, Russia;
}

\author{E. S. Andrianov}
\email{andrianov.es@mipt.ru}
\affiliation{
 Dukhov Research Institute of Automatics (VNIIA), 22 Sushchevskaya, Moscow 127055, Russia;
}
\affiliation{
 Moscow Institute of Physics and Technology, 9 Institutskiy pereulok, Dolgoprudny 141700, Moscow region, Russia;
}
\affiliation{
 Institute for Theoretical and Applied Electromagnetics, 13 Izhorskaya, Moscow 125412, Russia;
}
\date{\today}

\begin{abstract}
The dynamics of open quantum systems connected with several reservoirs attract great attention due to its importance in quantum optics, biology, quantum thermodynamics, transport phenomena, etc.
In many problems, the Born approximation is applicable which implies that the influence of the open quantum system on the reservoirs can be neglected.
However, in the case of a long-time dynamics or mesoscopic reservoir, the reverse influence can be crucial.
In this paper, we investigate the transient dynamics of several bosonic reservoirs connected through an open quantum system.
We use an adiabatic approach to study the temporal dynamics of temperatures of the reservoirs during relaxation to thermodynamic equilibrium.
We show that there are various types of temperature dynamics that strongly depend on the values of dissipative rates and initial temperatures.
We demonstrate that temperatures of the reservoirs can exhibit non-monotonic behavior.
Moreover, there are moments of time during which the reservoir with initially intermediate temperature becomes the hottest or coldest reservoir.
The obtained results pave the way for managing energy flows in mesoscale and nanoscale systems.

\end{abstract}

\maketitle

\section{Introduction}
Open quantum systems have gained a lot of attention over several decades.
A system can be treated as an open quantum system if it consists of a subsystem (qubit, molecule e.t.c.) that we are interested that interacts with its environment, e.g., photons of free space, phonons of host medium.
If the interaction between the subsystem and its environment is weak, one can exclude the environmental degrees of freedom from the consideration using Born approximation \cite{carmichael2009open,haake1973statistical,haake1969non} that assumes that the state of the environment is not changed due to interaction with the subsystem.
If one additionally uses Markov approximation that implies that the open quantum system dynamics is local in time \cite{carmichael2009open,haake1973statistical,haake1969non,louisell1973quantum,cohen1998atom}, one obtains master equation for the subsystem density matrix in Gorini-Kossakowsky-Sudarshan-Lindblad (GKSL) form \cite{lindblad1976generators, gorini1976completely, davies1974markovian}.
Different ratios between relaxation rates and open quantum system eigenfrequencies require different approaches for GKSL master equation application, namely, local approach \cite{shishkov2020perturbation, levy2014local, konopik2022local, LOC3}, global approach \cite{LOCvsGLB1, gonzalez2017testing, naseem2018thermodynamic, shishkov2019relaxation}, partial-secular approach \cite{cattaneo2019local, vovchenko2021model, EAS,vovcenko2023energy}, modified local approach \cite{trushechkin2021unified, potts2021thermodynamically}.
Some of these works have been dedicaded to the thermodynamics of the open quantum system and to the dynamics of the energy flows \cite{ LOC3, levy2014local, LOCvsGLB1, shishkov2019relaxation, vovcenko2023energy}.
These approaches have shown to be appropriate for the description of the dissipation in weak coupling limit.
Usually these are applied to the description of qubits \cite{reiter2013steady, leghtas2013stabilizing, martin2011dissipation, vovcenko2021dephasing}, quantum dots \cite{contreras2014dephasing}, molecules \cite{plenio2008dephasing}, nanostructures \cite{buvca2012note,flindt2004full,jonsson2017efficient}, nanolasers \cite{zyablovsky2017optimum, moy1999born, jonasson2016partially}.

The Born-Markov approximation assumes that the state of the reservoir - it is the Gibbs state with the given temperature - does not change \cite{schaller2014open, moy1999born, van1987born}.
This approximation is natural when one considers single reservoir with a large number of degrees of freedom.
However, there are many situation when the subsystem is connected with two reservoirs with different temperatures.
Such an approximation can not be valid, at least, long-time dynamics.
Indeed, in such a case the reservoirs interact through the subsystem, and their temperatures should get asymptotically equal.

The evolution of energy flows through the open quantum system and the evolution of the environment, especially when logical elements, diodes, or transistors are considered are very important in phononics and photonics \cite{li2012colloquium, chang2006solid, li2006negative, dubi2011colloquium} to prevent undesired energy flows.
Thus, an understanding of the reservoirs' temperatures dynamics at nanoscale are very important.

The direct solving (in particular, numerical) of the von Neumann equation for the reservoir is complicated and can be done only in the case of not too many modes (usually less than $100$) \cite{katz2008stochastic,baer1997quantum,gelman2003simulating,sergeev2023signature}.
Another possibility for solving this problem is to use the Zwanzig projection operator method to construct the master equation for the reservoir density matrix \cite{nakajima1958quantum, zwanzig1960ensemble, breuer2002theory}.
This approach implies solving complicated integral-differential equations.
One more option is to get equations on a number of quanta in the reservoir at some frequency under non-Markovian approaches, which are usually used in the limit of strong coupling with the reservoir.
This can be done via the Zwanzig projection operator method \cite{trushechkin2019dynamics, te2022understanding} or via Time Evolving Density Operator with Orthogonal Polynomials (TEDOPA mapping) \cite{nusseler2022fingerprint, chin2010exact, rosenbach2016efficient,tamascelli2019efficient}.
The first approach implies solving the integral-differential master equation for the open quantum system along with finding the number of quanta in a mode of the reservoir.
The second approach implies solving of a system of a large number of equations that describe a set of connected oscillators.
Both mentioned approaches do not take into account processes of transient dynamics of the reservoirs.
Because the terms of the reservoirs' Hamiltonians that are responsible for the transient dynamics usually do not directly take into account.
The absence of these terms leads to a non-Gibbs distribution of excitations in the reservoir \cite{borgonovi2016quantum,mori2018thermalization}.
However, if dynamics of an open quantum system and transient dynamics of reservoirs occur at different timescales \cite{vovchenko2021model, moy1999born}, it is possible to effectively consider transient dynamics.

In this paper, we consider an open quantum system coupled to $n$ bosonic reservoirs in the weak coupling limit.
We study the transient behavior of the reservoirs temperatures in the framework of developed adiabatic approximation.
In this approximation, we suppose the existence of two time scales.
At shorter time scale, the temperature of the reservoirs are fixed, and the open quantum system reaches its non-equilibrium stationary state with non-zero energy flow through it.
At longer time scale, the energy flow through the open quantum system results in a change in the reservoir temperatures.
Thus, each reservoir is in thermal equilibrium at shorter time scale, and their temperatures are changed due to the energy flow through the open quantum system at longer time scale.
We develop a general theory and apply it to the case of an oscillator-like open quantum system.
The interplay between dissipative rates and initial temperatures of reservoirs results in a consecutive equalization of reservoirs' temperatures during transient dynamics.
The last can result in the non-monotonical dynamics of reservoirs' temperatures.
If the dissipation rate into a reservoir with intermediate temperature is much smaller than the dissipation rates into the other reservoirs, this reservoir engages in temperature equalization in the last order.
Thus, it becomes the hottest one (or the coldest one) reservoir after some moment of time.
We calculate the thermal conductivity of the open quantum system in the case of two reservoirs and find the optimal ratio of the oscillator frequency and reservoir temperature for the fastest temperature equalization.

\section{The model}
We consider $n$ bosonic reservoirs describing by the Hamiltonians $\hat H_{Rj}, j=1,..,n$  interacting with a open quantum system (hereinafter, the open system) with Hamiltonian $\hat H_S$.
The total Hamiltonian reads
\begin{equation}\label{H}
\hat H=\hat H_S+\sum\limits_{j=1}^{n}\hat H_{Rj}+\sum\limits_{j=1}^{n}\hat H_{SRj},
\end{equation}
\begin{align}
{\hat H_{SRj}} = \epsilon_j {\hat S_{j}}{\hat R_{j}}.
\end{align}
Here $\hat S_{j}$ and  $\hat R_{j}$ are operators of the open system and the $j$-th reservoir respectiverly.
The operator $\hat H_{SRj} $ describes interaction between them with coupling constant $\epsilon_j$. 
In the Born-Markov approximation, the open system dynamics are governed by the Lindblad master equation \cite{lindblad1976generators, gorini1976completely, davies1974markovian, shishkov2019relaxation}
\begin{equation}\label{ME}
\frac{{\partial \hat \rho_S}}{{\partial t}} =  - i\left[ {{{\hat H}_S},\hat \rho_S } \right] + \sum_{j=1}^n \Lambda_j[\hat \rho_S] .
\end{equation}
where $\hbar=1$, $\Lambda_j \left[\hat \rho_S \right] $ is Lindblad superoperator describing interaction with $j$-th reservoir.
In the Born approximation, the reservoir state is constant during the evolution of the open system, and it is in thermal equilibrium with temperature $T_j$.
To this assumption be valid, the temperature and chemical potential of the reservoir should be constant at the time scale $ \tau_S$ --- characteristic dissipation time of the open system.
From this, it follows that $\tau_S$ should be much greater than $\tau_{Rj}$ --- characteristic time of $j$-th reservoir thermalization.
Further more, $\tau_S $ should be much smaller than $t_{\rm{eq}}$ ---  characteristic time of the temperature equilibration between the different reservoirs due to energy flows $J_j $ through the open system.
This conditions are fulfilled, if the reservoirs are large compared to the open system.
This means that the reservoirs should have a large number of modes.

The energy exchange between the open system and reservoirs happens when the open system transits from one eigenstate to another.
The quantum with the frequency that is equal to the difference of the mentioned eigenstate's eigenfrequencies is consumed by the reservoir.
This means, that the energy exchange between open system and reservoir happens only at some set of frequencies.
However, as $\tau_S$ is much larger than the time of establishment of Gibbs distribution in the reservoir, the energy consumed by the reservoir at some frequency is distributed among other energy levels of the reservoirs, and its Gibbs distribution sets up.

Thus, for the reservoir to be considered infinite number of modes, it is sufficient that the density of states of the reservoir be dense near transition frequencies of the open system.
The characteristic difference between eigenfrequencies of the reservoir is of the order $\Delta \nu= c/L$, where $L$ is the characteristical size of the reservoirs.
If the open system has the minimal transition frequency $\omega$, then mentioned condition can be written as $\Delta \nu/\omega=(c/L)/\omega\ll 1$.

Using Eq.~(\ref{ME}), the total energy flow into the open system can be defined as follows \cite{LOCvsGLB1,shishkov2019relaxation,vovcenko2023energy}
\begin{gather}\label{EnFlow0}
	\left< \dot{\hat H}_S \right> =\frac{d}{dt}\langle\hat \rho_S\hat H_S\rangle=\langle\dot{\hat{\rho}}_S\hat H_S\rangle= 
	\\ \nonumber
	= \sum\limits_{j=1}^n \mathrm{tr}(\Lambda_{j}[\hat \rho_S]\hat H_S)  \equiv \sum\limits_{j=1}^n J_j,
\end{gather}
where $J_j=\mathrm{tr}(\Lambda_{j}[\hat \rho_S]\hat H_S) $ is energy flow from the open system to the $j$-th reservoir.
Because the temperatures of reservoirs are constant on the time scale $\tau_S$, the energy flows $J_j $ are the functions of reservoirs' temperatures $ T_j$, $J_j = J_j \left( T_1,..., T_n, t\right)$ (in this work we are focused on the case of reservoirs with fixed volumes and zero chemical potentials).
On this time scale the density matrix tends to the stationary state for given temperatures of the reservoirs, $\hat{\rho}_S=\overline{\hat{\rho}}_S\left( T_1,..., T_n \right)$ (the overline denotes the stationaty value of the overlined expression).
Thus $\overline{\left< \dot{\hat H}_S \right>}=\sum_j \overline{J}_j=0$, here $\overline{J}_j=\mathrm{tr}(\Lambda_{j}[\overline{\hat \rho}_S]\hat H_S) $.
This means, that the energy of each reservoir is conserved on this timescale.
In turn, on time scale $t_{\rm{eq}}$, the energy flow from {\it{j}}-th reservoir to the open system changes the energy of the reservoir $E_j $
\begin{equation}\label{dE0}
	\frac{dE_j}{dt}=-\overline{J}_j \left( T_1,..., T_n \right).
\end{equation} 
Thus, $dE_j=(dE_j/dT_j)|_{T_j} dT_j = C_j(T_j)dT_j$ where $C_j(T_j)$ is the heat capacity of the $j$-th reservoir.
We assume that the heat capacity is the known function on $T$, as the reservoir is considered to be in thermal equilibrium under Born-Markov approximation.
Thus, we get the equations on the reservors temperatures dynamics
\begin{equation}\label{T_temp}
	\frac{dT_j}{dt}=-\overline{J}_j \left( T_1,..., T_n \right)/C(T_j).
\end{equation}

The stationary solution to these equations implies that all $J_j$ should be zero.
Generally, the equilibrium temperatures, as well as their time dependencies, strongly depend on the type of interaction of the open system with each reservoir.
In the subsequent section, we investigate time dynamics of $T_j $ for the case of harmonic oscillator interacting with reservoirs consisting of bosonic quasiparticles.

\section{Case of reservoirs of bosonic quasiparticles}

In this section, we apply the general equations~(\ref{T_temp}) for the case of $n$ bosonic reservoirs interacting with harmonic oscillator.
In this case, the Hamiltonians of the open  system and the bosonic reservoirs have the form
\begin{equation}\label{1Os_HS}
\hat H_S =  \omega \hat a^\dag \hat a,
\end{equation}
\begin{equation}
	{\hat H_{Rj}} = \sum\limits_k { {\omega _{kj}}\hat e_{kj}^\dag {{\hat e}_{kj}}},
\end{equation}
\begin{equation}\label{1Os_HSR}
{\hat H_{SRj}} =  \sum\limits_k {{\chi _{kj}}} (\hat a^\dag  + {{\hat a}})(\hat e_{kj}^ \dag  + {{\hat e}_{kj}}).
\end{equation}
Here, $\omega$ is oscillator frequency, $\hat a$ is annihilation operator for the oscillator, 
reservoirs are represented as sets of oscillators, $\hat e_{kj}$ is annihilation operator of the $k$-th oscillator in the $j$-th reservoior, $\omega_{kj}$ is frequency of the $k$-th oscillator in the $j$-th reservoior,
coefficients $\chi_{kj}$ describe the strength of interraction between oscillator and $k$-th oscillator in the $j$-th reservoior.
In the weak coupling regime between open system and reservoirs $\chi _{kj}\ll\omega$.
For such case, in the Born-Markov approximation (the reservoirs are in the Gibbs state with given temperatures $T_j$), one can obtain the master equation in GKSL form~(Eq.(\ref{ME})) with the following Lindblad superoperators \cite{vovchenko2021model}
\begin{equation}\label{SupOper}
	\Lambda_j[\hat \rho_S] =\frac{{{G_j}( - {\omega }, T_j)}}{2}\hat L[\hat a,\hat a^\dag] +\frac{{{G_j}({\omega},T_j)}}{2}\hat L[\hat a^\dag,\hat a].
\end{equation}
Here $\hat L [\hat X, \hat Y] = 2\hat X \hat \rho_S \hat Y-\hat Y \hat X \hat\rho_S -\hat\rho_S \hat Y \hat X$.
The coefficients ${G_j}(\pm\omega,T_j)$ determine the rates of transitions between eigenstates of the harmonic oscillator due to its interaction with $j$-th reservoir.
For the considered case, Eqs.~(\ref{1Os_HS}) -- (\ref{1Os_HSR}), they equal
\begin{equation}
	{G_j}(\pm\omega,T_j)=\gamma_j(\omega)(n_j(\omega,T_j)+1/2\mp1/2).
\end{equation}
Here
\begin{equation}
	n_j(\omega,T_j)=1/(\exp (\omega/T_j)-1),
\end{equation}
it is the mean occupancy of the $j$-th reservoir's states with eigenfrequency $\omega$, and
\begin{equation}
	\gamma_j\left(\omega\right) = \pi g_j\left(\omega\right) |\chi_j\left(\omega\right)|^2,
\end{equation}
where $g_j\left(\omega\right)$ is the $j$-th reservoir's density of states, $j=1,\ldots,n$.
The first term in Eq. (\ref{SupOper}) is responsible for the downward transitions and the second term in Eq. (\ref{SupOper}) is responsible for the upward transitions in the oscillator.
Note, that the Gibbs state of the reservoirs results in fulfillment of Kubo-Martin-Schwhinger condition, namely, $G_j(\omega)/G_j(-\omega)=e^{-\omega/T_j}$~\cite{breuer2002theory}.

The equations for the dynamics of the mean number of quanta can be found through the identity $d\langle \hat A\rangle/dt=\rm{tr({ \dot{\hat {\rho}}_S } \hat A)}$.
Using commutation relation $[\hat a,\hat a^\dag ]=\hat 1$ we obtain the following equation for the mean number of quanta $\left< \hat a^{\dag} \hat a \right> $ in oscillator:
\begin{align}
\frac{{\partial\langle {{\hat a^\dag}\hat a} \rangle }}{{\partial t}} =  
&\sum\limits_{j=1}^{n}- {G_j}( - \omega,T_j )\langle {{a^\dag }a}\rangle  + {G_j}(\omega,T_j )\left( {1 + \langle {{a^\dag }a}\rangle } \right).
\end{align}
The stationary solution to this equation is
\begin{gather}\label{aa_stat0}
\overline {\langle {{a^\dag }a}\rangle }  = \frac{{\sum\limits_{j=1}^{n}{G_j}(\omega,T_j ) }}{{ \sum\limits_{j=1}^{n}\left( G_j(-\omega, T_j) - G_j(\omega, T_j) \right) }}.
\end{gather}

Using the notations $\gamma_j(\omega)=\gamma_j$, $n_j(\omega, T_j)=n_j(T_j)$,~Eq.(\ref{aa_stat0}) can be rewritten in the form 
\begin{gather}\label{aa_stat}
\overline {\langle {{a^\dag }a}\rangle }  = \frac{\sum_{j=1}^n \gamma_j n_j(T_j) }{\sum_{j=1}^n \gamma_j}.
\end{gather}

From general expression~(\ref{EnFlow0}), the stationary energy flow, $\overline{J}_j$, is found to be
\begin{gather}\label{EnFlow}
\overline{J}_j=\omega ((G_j(\omega,T_j) - G_j(-\omega,T_j))\overline{\langle {{a^\dag }a}\rangle}  + {G_j}(\omega,T_j ))= 
\\ \nonumber
=\omega \frac{ \sum\limits_{k=1}^n \gamma_j \gamma_k (n_j(T_j) -n _k(T_k)) }{\sum\limits_{k=1}^n \gamma_k}=\\ \nonumber
=\omega \gamma_j \left( n_j (T_j) -  \sum\limits_{k=1}^n p_k n_k(T_k)\right).
\end{gather}
Here $p_k=\gamma_k/\sum_{m=1}^n \gamma_m$. 
In other words, the absolute value of the energy flow in $j$-th reservoir is defined by the difference between the $n_j$, occupancy of the state with frequency $\omega$ in the $j$-th reservoir, and mean occupancy of the state with frequency $\omega$ among all reservoirs. 
According to Eq.~(\ref{EnFlow}), normalized dissipation rates per unit time, $p_j$ can be interpreted as probabilities of absorption of the energy quantum $\omega $ by the $j$-th reservoir. 

Now, using the dependence of the stationary energy flows on the reservoir temperatures, we can apply Eqs.~(\ref{T_temp}) to determine the dynamics of the reservoirs' temperatures.
For the $n$ $D$-dimensional bosonic reservoirs with zero chemical potential, spectrum $\varepsilon(p)=cp^d$, and density of states $g(\varepsilon)=S_{D}\varepsilon^{D/d-1}/(2\pi\hbar)^D$, where $S_{D}=D\pi^{D/2}/\Gamma(D/2+1)$ is the surface area of unit sphere, the energy of each reservoir reads~\cite{landau2013statistical}
\begin{gather}
E=\int_\Gamma \frac{d^D x d^D p}{(2\pi \hbar)^D}\frac{\varepsilon}{\exp(\varepsilon/T)-1}=A V T^{\alpha+1}.
\end{gather}
Here $ A = S_{D} \Gamma(\alpha+1) \zeta(\alpha+1)/(2\pi\hbar)^D$ is a constant ($\zeta $is the Riemann zeta function), $\alpha=D/d$,  and $V$ is the volume of the reservoir. 
Thus, Eq. (\ref{dE0}) takes the form
\begin{equation}\label{dEdt}
	\frac{dE_j}{dt}=(\alpha_j+1)A_jV_jT^{\alpha_j}_j\frac{dT_j}{dt}=C_j(T_j) \frac{dT_j}{dt}=-\overline{J}_j,
\end{equation}
or, alternatively,
\begin{equation}\label{dEdt1}
	\frac{dT_j}{dt} = - \frac{\omega \gamma_j \left( n_j (T_j) -  \sum\limits_{k=1}^n p_k n_k(T_k)\right)}{(\alpha_j+1)A_jV_jT^{\alpha_j}_j}.
\end{equation}

The stationary state of this equation implies that $\overline{J}_j=0$ for each $j$.
Consequently, the equilibrium temperatures of reservoirs are determined by the following expression:
\begin{gather}\label{EqTemp}
	n_j(T_j^{{\rm{eq}}})=\sum_k p_k n_k(T_k^{{\rm{eq}}}).
\end{gather}
Right hand side of this equation is the same for every $j$ in the left hand side.
Thus, $n_j(T_j^{{\rm{eq}}})=n_k(T_k^{{\rm{eq}}})$, $\forall j,k \in \{1,\ldots,n\}$.
From this and the fact that $\partial n_j(T_j)/\partial T_j >0$, it follows that all temperatures of reservoirs should be the same in the stationary state: $T_j^{\mathrm{eq}}=T_{\mathrm{eq}}$.

The equilibrium temperature can be found by solving Eq. (\ref{EqTemp}) considering $T_j^{\mathrm{eq}}=T_{\mathrm{eq}}$.
However, a more easy way is to establish some integrals of motion in the Eq. (\ref{dEdt1}).
From Eq. (\ref{dEdt1}) and the definition of $p_k$ it follows that total energy of reservoirs is conserved
\begin{equation}
	\sum_j\frac{dE_j}{dt}=\sum_j(\alpha_j+1)A_jV_jT^{\alpha_j}_j\frac{dT_j}{dt}=0.
\end{equation}

Thus, the equilibrium temperatures can be found from total energy conservation
\begin{equation}\label{Teq}
	\sum\limits_{j=1}^{n} A_j V_{j}T_j^{\alpha_j+1}=\sum\limits_{j=1}^{n} A_j V_{j}T_{eq}^{\alpha_j+1}.
\end{equation}
Note, that from $\partial E_j (T_j)/\partial T_j>0$ it follows that Eq. (\ref{Teq}) has only one positive real solution $T_{eq}$.

From Eq. (\ref{EnFlow}) one can conclude that if $n_j(T_j)$ is greater than mean occupancy $n_j(T_j)=\sum_k p_k n_k(T_k)$, the reservoir is cooling down, and if $n_j(T_j)$ is smaller than mean occupancy, the reservoir is heating up.
Because $\partial n_j(T_j)/\partial T_j >0$, for $T_m=\max T_j$  one has $n_m(T_m) > n_j(T_j)$.
Thus, $n_m(T_m) > \sum_k p_k n_k(T_k)$, i.e., $n_m$ maximal number from a set which is greater than the average of the set.
Thus, in all moments of time the hottest reservoir is cooling.
Analogously, the coldest reservoir is heating up at all moments of time.

Let's consider a function that represents the maximal instant temperature difference between the reservoirs $f(t)=\max_j T_j(t) - \min_j T_j(t)$.
By definition, the $f(t)\ge 0$.
Also, because the coldest reservoir is heating up and the hottest reservoir is cooling up, $f(t)$ is a decreasing function in time.
If we suppose that $f(t)\rightarrow\delta>0$ at $t \rightarrow \infty$, then $\max_j n_j(T_j)>\sum_k p_k n_k(T_k)>\min_j n_j(T_j)$ when $t\rightarrow \infty$.
Thus the energy flows in the hottest reservoir and the coldest reservoir tend to a finite value when $t\rightarrow \infty$.
This means that the heat capacities of these reservoirs should tend to the infinity at some temperature.
As we consider the case of finite heat capacities, it follows that $f(t)\rightarrow 0$.
Consequently, at the stationary states, all reservoirs' temperatures are the same.
From the arguments presented above it follows that the stationary states of the reservoirs with the temperature determined by Eq.~(\ref{EqTemp}) are stable with respect to small perturbation.

\section{Transient dynamics of reservoirs' temperatures}
\subsection{Evaluation of time of temperature equalization}

Let us consider the case of two reservoirs.
In this case, Eqs. (\ref{aa_stat}) -- (\ref{EnFlow}) are reduced to
\begin{equation}\label{Mean_aa}
\overline {\langle {{a^\dag }a}\rangle }  = \frac{{{G_1}(\omega, T_1 ) + {G_2}(\omega, T_2 )}}{{ G_1(-\omega, T_1) - G_1(\omega, T_1) + G_2(-\omega, T_2) -G_2(\omega, T_2) }},
\end{equation}
\begin{gather}\label{J1}
\overline{J}_1=\omega \Gamma_\omega (n_1(\omega, T_1)-n_2(\omega,T_2)),
\end{gather}
where $\Gamma_\omega=\gamma_1 \gamma_2 /(\gamma_1 + \gamma_2)$.

It is possible to define the energy flow through the equality $\bar J = - \kappa \Delta T$, where $\kappa\left(T_1,T_2\right) $ is the thermal conductivity and $\Delta T= T_2-T_1$.
A characteristic time of the establishment of thermal equilibrium is $t_{{\rm{eq}}} = C(T)/\kappa $.
Using Eqs.~(\ref{Mean_aa}) and~(\ref{J1}), the thermal conductivity $\kappa\left(T_1,T_2\right) $ is found to be
\begin{gather}
\kappa(T_1,T_2)=- \frac{\omega\Gamma_\omega}{ T_1 - T_2}   \frac{e^{\omega /T_1}-e^{\omega /T_2}}{{{{\left( {{e^{\omega /T_1}} - 1} \right)\left( {{e^{\omega /T_2}} - 1} \right)}} }}=\\ \nonumber
=- \frac{2\omega\Gamma_\omega}{ T_1 - T_2}\frac{\mathrm{sh}\left(\cfrac{\omega}{2T_1}-\cfrac{\omega}{2T_2}\right)}{\mathrm{sh}\left(\cfrac{\omega}{2T_1}\right)\mathrm{sh}\left(\cfrac{\omega}{2T_2}\right)}. 
\end{gather}
If $\Delta T=T_2-T_1\ll T_1 \equiv T$, then
\begin{gather}
{{\bar J}_1} =  \omega \Gamma_\omega  (n(T)-n(T+\Delta T))=-\omega \Gamma_\omega\frac{\partial n(T)}{\partial T} \Delta T  \\ \nonumber
= -  \Gamma_\omega   \frac{{{ e^{\omega /T}}\omega^2 /{T^2}}}{{{{\left( {{e^{\omega /T}} - 1} \right)}^2} }} \Delta T
= -  \Gamma_\omega   \frac{{(\omega /{2T})^2}}{\mathrm{sh}^2(\omega/2T)} \Delta T.\\ \nonumber
\end{gather}
Thus, the thermal conductivity $\kappa\left(T_1,T_2\right) = \kappa\left(T\right) $ equals
\begin{equation}
\kappa(T)= \Gamma_\omega   \frac{{{ e^{\omega /T}}\omega^2 /{T^2}}}{{{{\left( {{e^{\omega /T}} - 1} \right)}^2} }}=\Gamma_\omega   \frac{{(\omega /{2T})^2}}{\mathrm{sh}^2(\omega/2T)}.
\end{equation}
For $\omega/T\gg 1$ we have $\kappa(T)\approx  \Gamma_\omega (\omega^2/T^2)e^{-\omega/T}$, while for $\omega/T\ll 1$ we obtain $\kappa(T)\approx  \Gamma_\omega$.

Consequently, the characteristic time of temperature equalization between two reservoirs with $|\Delta T (0)|/T_{1,2}(0)\ll 1$ can be  estimated as
\begin{gather}\label{t_ch}
t_{{\rm{eq}}}=\frac{C_{min}(T)}{\kappa(T)}=\frac{C_{min}(T)}{\Gamma_\omega}\frac{\mathrm{sh}^2(\omega/2T)}{(\omega/2T)^2}.
\end{gather}
where $C_{min}(T)=\mathrm{min}(C_1(T),C_2(T))$. 

For $\omega/T\gg 1$, the characteristic time of temperature equalization can be evaluated as $t_{{\rm{eq}}} \simeq (C_{min}(T)/\Gamma_\omega)(e^{\omega/T}/(\omega/T)^2)$, and thus, $t_{{\rm{eq}}} $ increases with increase of the ratio $\omega/T $.
For $\omega/T\ll 1$, we have $t_{{\rm{eq}}} \simeq C_{min}(T)/\Gamma_\omega$ and $t_{{\rm{eq}}} $ increases with decrease of the ratio $\omega/T $.
Thus, there are the value $\omega/T$ that minimizes the time of temperature equalization determined by the equation
\begin{gather}\label{tmin}
\frac{\omega/T}{\alpha_{\mathrm{min}}+2}=\mathrm{th}\left(\frac{\omega}{2T}\right).
\end{gather}
where $\alpha_{\mathrm{min}}$ equals $\alpha$ of the reservoir with minimal heat capacity.

\begin{figure}[h]
\center{\includegraphics[width=0.95\linewidth]{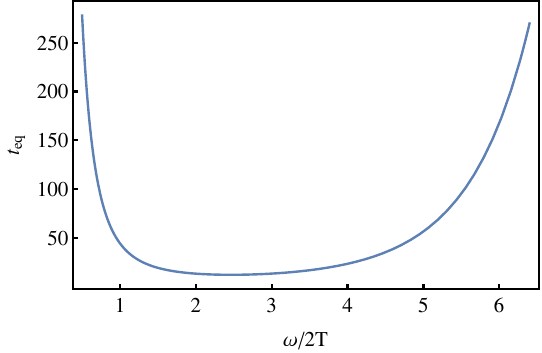}}
\caption{The dependence of $t_{{\rm{eq}}}$ on the ratio $\omega/2T$
Parameters of the reservoirs are set $A_1=A_2=1$, $V_1=1$, $V_2=1$, $\alpha_1=\alpha_2=3$. }
\label{t_eq}
\end{figure}

The dependence of $t_{{\rm{eq}}} $ on the ratio $\omega/T $ is shown on Fig. \ref{t_eq}.
It is a function with the minimum at the ratio $\omega/T \simeq 5$ and a wide "flat" range near this minimum.

Note, that $\tau_S\sim \Gamma_\omega ^{-1}$. 
For the model to be applicable the condition $t_{{\rm{eq}}}/\tau_S \gg 1$ should be satisfied.
This means that $C_{min}(T(t))\gg 1$.
For the considered reservoirs of quasiparticles with zero chemical potential, this condition means that occupation number of quasiparticles in the reservoir should be much greater than $1$.

\subsection{Variation of temperatures' sequence}
\begin{figure}[h]
\includegraphics[width=0.95\linewidth]{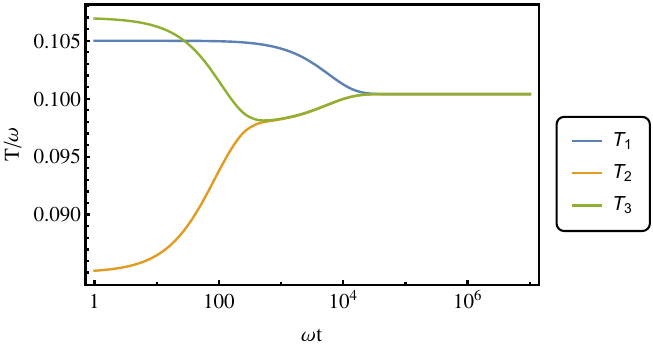}
\caption{The time dependence of temperature of the three reservoirs connected through the oscillator.
The parameters: $\omega=1$, $\alpha_j=3$, $\gamma_1=10^{-4}$, $\gamma_2=2\cdot 10^{-2}$, $\gamma_3=2\cdot 10^{-2}$, $A_j=1$, $V_j=1$.
Initial temperatures of the reservoirs are $T_1 = 0.105$, $T_2= 0.085$, $T_3 = 0.107$.}
\label{Hot_change}
\end{figure}

The developed model predicts that it is possible that hottest and coldest reservoirs cease to be so during evolution.
As has been mentioned, from the Eq. (\ref{EnFlow}) it follows that if $n_j(T_j)$ is greater than mean occupancy, the reservoir is cooling down, and if $n_j(T_j)$ is smaller than mean occupancy, the reservoir is heating up.
Thus, the coldest reservoir never can become the hottest reservoir and vice versa.
However, other pairs of reservoirs do not obey this restriction.
For example, if the dissipative rate of the open system in the first reservoir is much less than the dissipative rates of the open system in the second and third reservoirs then the second and third reservoirs become to have equal temperatures, and only after that, they become to have equal temperature with the first reservoir.
As a result, the temperature of the first reservoir, being initially intermediate, becomes the highest before temperature equalization with the second and third reservoirs, see Fig. \ref{Hot_change}.

\subsection{Transient non-monotonic temperature behavior}
The developed theory also predicts the possibility of non-monotonic evolution of reservoirs' temperatures.
For example, this is possible in the following case.
Different pairs of reservoirs can have different $t_{{\rm{eq}}}$ time.
As we consider $n$ reservoirs, then we have $n(n-1)/2$ values of $t_{{\rm{eq}}}$ times.
The set of reservoirs, that have minimal value of time of temperature equalization, $t_{{\rm{eq}}}$, among these $n(n-1)/2$ $t_{{\rm{eq}}}$ values, their temperatures become equal first.
Suppose that there are a subset of $m$ reservoirs in this set.
All of these $m$ reservoirs have the same temperature after their temperature become equal to each other, and can be considered as single reservoir with effective heat capacity.
This reduces the number of reservoirs in the system.
Now we have $(n-m+1)(n-m)/2$ values of $t_{{\rm{eq}}}$ times, and the process repeats until only one reservoir is left.
Thus, temperature equalization between all reservoirs can be divided into consequent temperature equalization of reservoirs in each subsets.

\begin{figure}
\includegraphics[width=0.95\linewidth]{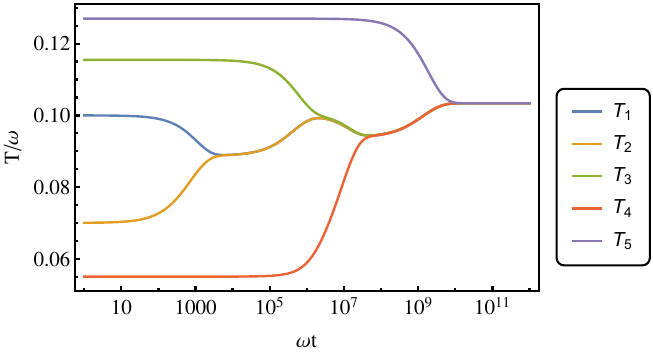}
\caption{
The time dependence of temperatures of the five reservoirs connected through the oscillator. 
The parameters: $\omega=1$, $\alpha_j=3$, $\gamma_1=10^{-2}$, $\gamma_2= 10^{-3}$, $\gamma_3=10^{-6}$, $\gamma_4=10^{-7}$, $\gamma_5=3\cdot 10^{-9}$, $A_j=1$, $V_j=1$. Initial temperatures of the reservoirs are $T_1 = 0.1$, $T_2= 0.07$, $T_3 = 0.1155$, $T_4 = 0.055$, $T_5 = 0.127$.}
\label{Oscill}
\end{figure}

In the case when initial subsets of reservoirs have sufficient differences in initial temperatures and $t_{{\rm{eq}}}$, non-monotonic dynamics of temperatures are possible.
An example of such behavior is represented in Fig. \ref{Oscill} (see $T_1$).
Here, the first and the second reservoirs become to have equal temperatures first.
The initial temperature of the first reservoir is higher than the initial temperature of the second reservoir.
As a result, the temperature of the first reservoir decreases.

After the first and the second reservoirs become to have equal temperatures, they both become to have equal temperature with the third reservoir that has a greater initial temperature than they both.
Thus, the equal temperature of these three reservoirs is greater than the equal temperature of the first two reservoirs.
As a consequence, the temperature of the first reservoir, first, decreases and, the, increases.
Another words, non-monotonic temperature's dynamics take place.

\section{Conclusion}
In this work, we considered an open quantum system connected to several reservoirs in the weak coupling regime.
We studied the transient temperature regimes of reservoirs during temperature equalization via the open quantum system.
We showed that the interplay between dissipative rates and occupancies of reservoirs results in various transitional temperature regimes.
We showed, that it is possible to achieve non-monotonic dynamics of the reservoir temperature by right choosing the initial temperatures and dissipative rates of the reservoirs.
Moreover, by lowering the dissipative rate of one reservoir, one can increase the time that is needed for this reservoir to equalize temperature with other ones.
This can make a reservoir with intermediate temperature the hottest or the coldest one after some moment of time.

For the oscillator-like open quantum system, the explicit formula for the stationary energy flow from a reservoir to the open quantum system that reveals the interplay between dissipative rates and occupancies of reservoirs was derived.
The energy flow into the $j$-th reservoir was shown to be proportional to the difference between the occupancy of the $j$-th reservoir at the eigenfrequency of the open quantum system and the mean occupancy at this eigenfrequency among all reservoirs.
The mean occupancy is equal to the sum of the reservoirs' occupancies weightened with normalized dissipative rates.

In the case of two reservoirs, we calculated the thermal conductivity of the open quantum system, with the help of which we estimated the time needed for the temperature equalization of these reservoirs.
Moreover, we showed that there exists an optimal ratio of the transitional frequency and reservoir temperature that allows to minimize the time of the temperature equalization between reservoirs.
\\

\section*{Acknowledgement}
This work was financially supported by the Russian Science Foundation (project no. 20-72-10057).
I.V.V. acknowledges the support of the Foundation for the Advancement of Theoretical Physics and Mathematics BASIS(grant \#22-1-5-55-1).

\bibliography{Thermal_conductivity_u0}

\end{document}